\documentclass[prd,twocolumn,superscriptaddress,floatfix,nofootinbib]{revtex4}
\usepackage{amsfonts}
\usepackage[dvipdfm,colorlinks=true, citecolor=blue, linkcolor=blue, urlcolor=blue]{hyperref}
\usepackage{color,xcolor,graphicx,amsfonts,multirow,amsmath,multirow}
\allowdisplaybreaks

\begin{document}

\title{$B_c \to J/\psi$ helicity form factors and the $B_c^+ \to J/\psi+(P, V, \ell^+ \nu_\ell)$ decays}

\author{Wei Cheng}
\address{School of Science, Chongqing University of Posts and Telecommunications, Chongqing 400065, P.R. China}
\address{State Key Laboratory of Theoretical Physics, Institute of Theoretical Physics, Chinese Academy of Sciences, Beijing, 100190, P.R. China}
\author{Yi Zhang}
\address{Department of Physics, Chongqing University, Chongqing 401331, P.R. China}
\author{Long Zeng}
\address{Department of Physics, Chongqing University, Chongqing 401331, P.R. China}
\author{Hai-Bing Fu\footnote{Corresponding author}}
\email{fuhb@cqu.edu.cn}
\address{Department of Physics, Guizhou Minzu University, Guiyang 550025, P.R. China}
\address{Chongqing Key Laboratory for Strongly Coupled Physics, Chongqing 401331, P.R. China}
\author{Xing-Gang Wu}
\email{wuxg@cqu.edu.cn}
\address{Department of Physics, Chongqing University, Chongqing 401331, P.R. China}
\address{Chongqing Key Laboratory for Strongly Coupled Physics, Chongqing 401331, P.R. China}
\date{\today}

\begin{abstract}
In this paper, we calculate the $B_c\to J/\psi$ helicity form factors (HFFs) up to twist-4 accuracy by using the light-cone sum rules (LCSR) approach. After extrapolating those HFFs to the physically allowable $q^2$ region, we investigate the $B^+_c$-meson two-body decays and semi-leptonic decays $B_c^+ \to J/\psi+(P, V, \ell^+ \nu_\ell)$ with $P/V$ stands for light pseudoscalar/vector meson, respectively. The branching fractions can be derived by using the CKM matrix element and the $B_c$ lifetime from the Particle Data Group, and we obtain ${\cal B}(B_c^+ \to J/\psi \pi^+)=(0.136^{+0.002}_{-0.002})\%$, ${\cal B}(B_c^+ \to J/\psi K^+)=(0.010^{+0.000}_{-0.000})\%$, ${\cal B}(B_c^+ \to J/\psi \rho^+) =(0.768^{+0.029}_{-0.033})\%$, ${\cal B}(B_c^+ \to J/\psi K^{\ast +})=(0.043^{+0.001}_{-0.001})\%$, ${\cal B}(B_c^+ \to J/\psi \mu^+\nu_\mu)=(2.802^{+0.526}_{-0.675})\%$ and ${\cal B}(B_c^+ \to J/\psi \tau^+\nu_\tau)=(0.559^{+0.131}_{-0.170})\%$. We then obtain ${\cal R}_{\pi^+/\mu^+\nu_\mu} = 0.048^{+ 0.009}_{-0.012}$ and ${\cal R}_{K^+ / \pi^+} = 0.075^{+0.005}_{-0.005}$, which agree with the LHCb measured value within $1\sigma$-error. We also obtain ${\cal R}_{J/\psi}=0.199^{+ 0.060}_{-0.077}$, which like other theoretical predictions, is consistent with the LHCb measured value within $2\sigma$-error. Those imply that the HFFs under the LCSR approach are also applicable to the $B^+_c$ meson two-body decays and semi-leptonic decays $B_c^+ \to J/\psi+(P, V, \ell^+ \nu_\ell)$, and the HFFs obtained by using LCSR in a new way implies that there may be new physics in the $B_c\to J/\psi \ell^+ \nu_\ell$ semi-leptonic decays.
\end{abstract}


\maketitle

\section{Introduction}
The $B_c$-meson that consists of two different heavy quarks was first observed by CDF collaboration~\cite{CDF} in year 1998. Since then, the properties of the $B_c$-meson have been?studied?extensively, cf. the reviews~\cite{Brambilla:2010cs, Brambilla:2004wf, Li:2018lxi}. Its related decay processes can provide good platforms not only to understand the non-perturbation interactions by fixing the CKM-matrix elements, but also to further searches for new physics beyond the standard model. In recent years, many ratios of different branching fraction associated with $B_c \to J/\psi$ have been measured by the LHCb collaboration, such as ${\cal R}_{\pi^+/\mu^+\nu_\mu}= {\cal B}(B_c^+ \to J/\psi \pi^+)/{\cal B}(B_c^+ \to J/\psi \mu^+ \nu_\mu)= 0.0469\pm 0.0028({\rm stat.})\pm 0.0046({\rm syst.})$~\cite{Aaij:2014jxa}, ${\cal R}_{K^+/\pi^+}={\cal B}(B_c^+ \to J/\psi K^+)/{\cal B}(B_c^+ \to J/\psi \pi^+)=  0.069 \pm 0.019({\rm stat.})\pm 0.005({\rm syst.})$~\cite{Aaij:2013vcx}, which is updated to ${\cal R}_{K^+ / \pi^+}=0.079 \pm 0.007 ({\rm stat.}) \pm 0.003 ({\rm syst.})$~\cite{Aaij:2016tcz}, and ${\cal R}_{J/\psi} = {\cal B}(B_c^+ \to J/\psi \tau^+ \nu_\tau)/{\cal B}(B_c^+ \to J/\psi \mu^+ \nu_\mu)= 0.71\pm0.17(\rm stat.)\pm 0.18(\rm syst.)$~\cite{Aaij:2017tyk}, where ``stat." is the statistical error and ``syst." is the systematic error.

Many theoretical studies on those decay channels have been published in the literature. For the branching ratio ${\cal R}_{\pi /\mu^+ \nu_\mu}$, theoretical predictions of Refs.\cite{Ke:2013yka, AbdElHady:1999xh, Issadykov:2018myx} agree with the measured value of Ref.~\cite{Aaij:2014jxa} within the $2\sigma$-error and the predictions of Refs.\cite{Ebert:2003cn, Chang:2014jca} are within $1\sigma$-error. For the branching ratio ${\cal R}_{K^+ / \pi^+}$, the predictions of Refs.\cite{Ebert:2003cn, AbdElHady:1999xh, Issadykov:2018myx, Chang:2014jca, Wen-Fei:2013uea} agree with the latest LHCb measurement~\cite{Aaij:2016tcz} within $1 \sigma$-error. For the branching ratio ${\cal R}_{J/\psi}$, the predictions of Refs.\cite{Leljak:2019eyw, Issadykov:2018myx, Wen-Fei:2013uea, Azizi:2019aaf} agree with LHCb measured value~\cite{Aaij:2017tyk} within $2 \sigma$-error. The key components and main error sources of those decays are $B_c \to J/\psi$ transition form factors (TFFs), which have been studied under various approaches, such as the relativistic model (RM)~\cite{AbdElHady:1999xh}, the light-front constituent quark model (LFQM)~\cite{Ke:2013yka}, the covariant confined quark model (CCQM)~\cite{Issadykov:2018myx}, the Bethe-Salpeter model (BSM)~\cite{Chang:1992pt}, the Bethe-Salpeter relativistic quark model (BS RQM)~\cite{Chang:2014jca}, the relativistic quasi-potential Schr\"odinger model (RQM)~\cite{Ebert:2003cn}, perturbation Quantum Chromodynamics (pQCD)~\cite{Wen-Fei:2013uea}, the light-cone sum rule (LCSR)~\cite{Leljak:2019eyw}, Quantum Chromodynamics sum rules (QCDSR)~\cite{Azizi:2019aaf}, and the covariant light-front constituent quark model (CLFQM)~\cite{Huang:2018nnq}, etc. There still exist a large discrepancy between different approaches, which attract the theorists to inquire further.

In the present paper, we shall study the $B_c \to J/\psi$ helicity form factors (HFFs) instead of traditional TFFs by applying the LCSR approach.
The main difference between?the two methods is the method of decomposition. The $\gamma$-structures of the hadronic matrix elements can be decomposed into Lorentz-invariant structures by using covariant decomposition for the TFFs decomposition method, while a new combination of Lorentz-invariant structures are obtained by using the polarization vectors of the off-shell W-Boson for the HFFs decomposition method. The use of HFFs has some advantages in comparison to the usual treatment of TFF~\cite{Bharucha:2010im, Fu:2014pba, Fu:2014uea, Fu:2016yzx, Fu:2018vap, Cheng:2018ouz, Fu:2020vqd}: I) The diagonalized unitarity relations will lead to the dispersion bounds that will be represented specifically by constraining the coefficients in the series expansion (SE) or simplified series expansion (SSE) parameterization; II) The polarized decay widths of a meson transition to a vector meson can be well studied by using the HFFs; III) As the helicity amplitude has a definite spin-parity quantum number, the helicity amplitude is convenient to consider the contribution from the excited state of meson in the transition process. After calculating the $B_c \to J/\psi$ HFFs within the LCSR, we shall extend them into allowable physical region and then investigate the properties of $B_c^+ \to J/\psi+(P, V,\ell^+\nu_\ell)$ decays.

The remaining parts of the paper are organized as follows. In Sec.~\ref{Section:II}, we present the calculation technology of the $B_c \to J/\psi$ HFFs and the branching fractions for $B_c^+ \to J/\psi+(P, V, \ell^+ \nu_\ell)$ decays. In Sec.~\ref{Section:III}, we give the numerical results and discussions. Section~\ref{Section:IV} is reserved for a summary.

\section{Theoretical Framework}\label{Section:II}

The $B^+_c$-meson two-body decays $B_c^+\to J/\psi+(P, V)$ are achieved predominantly via a $\bar b\to\bar c u \bar q$ transition, which can be expressed as the corresponding hadron matrix element $\langle J/\psi + P (V)|\mathcal{O}|B^+_c\rangle$.
Hrer, the non-factoring effect is mainly the exchange of hard gluons, so the contribution of this part can be dealt with by perturbation theory. Considering the heavy quark limit, the hadron matrix element under the leading order approximation of $\alpha_s$ can be expressed by the simple factorization formula. The corresponding transition amplitudes can be expressed as follows:
\begin{align}
&{\cal A}(B^+_c\!\!\to\! J/\psi P) \!=\!\frac{G_F\! V_{cb}\!V_{uq}^\dagger}{\sqrt 2}a_1\! \langle P|\bar{q}\gamma^\mu\gamma_5 u|0\rangle
\langle J/\psi|j_\mu |B^+_c\rangle,
\\
&{\cal A}(B^+_c\!\!\to\! J/\psi V)  \!=\frac{G_F V_{cb}V_{uq}^\dagger}{\sqrt 2}a_1\langle V|\bar{q}\gamma^\mu u|0\rangle
\langle J/\psi|j_\mu |B^+_c\rangle,
\end{align}
where $\langle P|\bar{q}\gamma^\mu\gamma_5 u|0\rangle=-i f_P q^\mu$ with $P$ being a pseudoscalar meson $\pi^+$ or $K^+ $, $\langle V|\bar{q}\gamma^\mu u|0\rangle= m_V f_V \epsilon_\mu^*$ with $V$ bing a vector meson $\rho^+$ or $K^{\ast +}$, and the hadronic matrix elements $\langle J/\psi| j_\mu |B^+_c\rangle$ with $j_\mu=\bar{c}\gamma_\mu(1-\gamma^5) b$ can be projected as the HFFs ${\cal H}_\sigma^{B_c\to J/\psi}(q^2)$, see Eq.~\eqref{HFF:Definition}. The $a_1=C_2+\xi C_1$ with $\xi=1/N_c$~\cite{Buchalla:1995vs}. Thus, the decay widths of $B_c^+ \to J/\psi +(P, V)$ in terms of the amplitudes ${\cal A}(B^+_c\to J/\psi  +(P, V))$ are given by
\begin{align}
\Gamma(B^+_c\to &J/\psi \!+\!(P, V)) \!=\!\frac{|{\bf p_2}||{\cal A}(B^+_c\to J/\psi +(P,V))|^2}{8\pi m^2_{B_c}} \nonumber \\
& =
\frac{G^2_F |{\mathbf p_2}|}{16 \pi m^2_{B_c}} \frac{\lambda(m_{(P,V)}^2)}{m_{(P,V)}^2}
\left|V_{cb}V_{uq}^\dagger a_1 f_{(P,V)} m_{(P, V)}\right|^2 \nonumber \\
&\times \sum\limits_{i}\left({\cal H}_i^{B_c\to J/\psi}(m^2_{(P,V)} )\right)^2,\label{Eq:DW}
\end{align}
where the fermi constant $G_F=1.166\times10^{-5}~{\rm GeV}^{-2}$ \cite{Zyla:2020zbs}, the momentum of the off-shell W-boson in the $B_c$ rest frame is $|{\bf p_2}|=\lambda^{1/2}/(2m_{B_c}^2)$ with the usual phase-space factor  $\lambda = (m_{B_c}^2 + m_{J/\psi}^2 - m_{(P,V)}^2)^2 -4m_{B_c}^2 m_{J/\psi}^2$. $f_{(P, V)}$ and $m_{(P, V)}$ are the decay constants and mass. For a pseudoscalar meson $P=\pi^+ (K^+)$, thus $q=d(s)$ and $i=t$, while, for a vector meson $V=\rho^+ (K^{\ast +})$, then $q=d (s)$ and $i=(0,1,2)$. The corresponding mass and decay constant are listed in Table.~\ref{tab:mass_DC}.

\begin{table}[t]
\caption{The mass and decay constant of $(P, V)$-mason in the $B_c$-meson two body decay.}\label{tab:mass_DC}
\centering
\begin{tabular}{lll}
\hline
~~~~~~~~~~~~&Mass~~~~~~~~~~~~~~~~~~~ & Decay constant\\
\hline
$\pi^+$    & $0.140~\rm{GeV}$~\cite{Tanabashi:2018oca}   &$0.1304~\rm{GeV}$~\cite{Khodjamirian:2009ys}\\
$K^+$      & $0.494~\rm{GeV}$~\cite{Tanabashi:2018oca}   &$0.1562~\rm{GeV}$~\cite{Agashe:2014kda}\\
$\rho^+$     & $0.775~\rm{GeV}$~\cite{Tanabashi:2018oca}   &$0.2210~\rm{GeV}$~\cite{Issadykov:2018myx}\\
$K^{*+}$   & $0.892~\rm{GeV}$~\cite{Tanabashi:2018oca}   &$0.2268~\rm{GeV}$~\cite{Issadykov:2018myx}\\
\hline
\end{tabular}
\end{table}
The amplitudes of $B^+_c$-meson semi-leptonic decays $B_c^+ \to J/\psi \ell^+ \nu_\ell$ can be factored into:
\begin{equation}
{\cal A}(B^+_c\!\to\! J/\psi \ell^+ \nu_\ell)=  \frac{G_F V_{cb}}{\sqrt 2 }
 \ell^+ \gamma^\mu(1-\gamma^5) \nu_\ell \langle J/\psi| j_\mu |B_c^+\rangle.
\end{equation}
After completing the square of the sime-leptonic decay amplitude~\cite{Zhou:2019stx}, the corresponding decay widths in terms of the HFFs ${\cal H}_\sigma^{B_c\to J/\psi}(q^2)$ can be written as
\begin{align}
\Gamma(B_c^+\to & J/\psi  \ell^+\nu_\ell)
=\frac{G_F^2|V_{cb}|^2}{12 (2\pi)^3} \int_{q^2_{\rm min}}^{q^2_{\rm max}} dq^2 \bigg( 1 - \frac{m_\ell^2}{q^2}\bigg)^2
\nonumber\\
& \times\lambda \frac{|{\bf p_2}|}{m_{B_c}^2} \bigg[\bigg(1+\frac{m_\ell^2}{2q^2}\bigg) \sum\limits_{i=0,1,2} ({\cal H}_i^{B_c\to J/\psi}(q^2))^2
\nonumber\\
& +\frac{3m_\ell^2}{2q^2} ({\cal H}_t^{B_c\to J/\psi}(q^2))^2\bigg],\label{Eq:Semi}
\end{align}
where $\ell$ stands for the lepton $\mu$ or $\tau$, the range of integration are from $q_{\rm min}^2=m_{\ell}^2$ to $q_{\rm max}^2=(m_{B_c}-m_{J/\psi})^2$.

Then, the key component of the decay width of $B_c \to J/\psi$ transitions are the HFFs ${\cal H}_\sigma^{B_c\to J/\psi}(q^2)$ with $\sigma=(0,1,2,t)$ defined in term of the hadronic matrix elements as follows:
\begin{align}
{\cal H}_{\sigma}^{B_c\to J/\psi} &(q^2)= \sqrt{\frac{q^2}{\lambda}} \sum\limits_{\alpha=0,\pm,t}{\varepsilon_\sigma^{*\mu}(q)}
\nonumber\\
&\times \langle J/\psi(k,\varepsilon_\alpha(k)) |\bar c \gamma_\mu(1-\gamma^5) b |B_c(p)\rangle,
\label{HFF:Definition}
\end{align}
where $\varepsilon_\alpha(k)$ with momentum $k=(k^0,0,0,|\vec{k}|)$ are $J/\psi$-meson longitudinal ($\alpha=0$) and transverse ($\pm$) polarization vectors, ${\varepsilon_\sigma (q)}$ with transfer momentum $q=(q^0,0,0,-|\vec q |)$ are the polarization vectors of the off-shell $W$-boson, and $\sigma=(0,1,2,t)$. These momenta satisfy the relationship $q=(p-k)$. HFFs is a novel decomposition method that multiplies the polarization vectors of the off-shell W-Boson (${\varepsilon_\sigma (q)}$) by the hadronic matrix element. Compared with TFFs, this factor refactorizes the hadron element by covariant decomposition instead of by the polarization vectors of the off-shell W-Boson (${\varepsilon_\sigma (q)}$). It can be proved that HFFs under this new decomposition method has Lorentzian-invariant property as well as TFFs. A simple proof can also be made directly from the expression of HFFs, since HFFs are composed of TFFs, and TFFs have Lorentzian-invariant properties, HFFs also have Lorentzian invariant properties. In conclusion, HFFs is a decomposition method as effective as the TFFs method.

Further, it can be found that the helicity amplitude consists of a single TFF or multiple TFFs mixed in the TFFs~\cite{Leljak:2019eyw}, while the helicity amplitude corresponds to HFFs in HFFs, and it can be computed directly in LCSR.

The calculated procedure of the $B_c\to J/\psi$ HFFs ${\cal H}_\sigma^{B_c\to J/\psi}(q^2)$ under the LCSR approach is the same as that in Ref.~\cite{Cheng:2018ouz,Fu:2020vqd}. We thus will directly give the LCSR results of $B_c\to J/\psi$ HFFs as follows:
\begin{widetext}
\begin{eqnarray}
{\cal H}_{0}^{B_c\to J/\psi}(q^2)
&=& \int_0^1 du e^{(m_{B_c}^2 - s)/M^2} \frac{m_b f_{J/\psi}^\bot {\cal Q}}{2\sqrt\lambda  m_{J/\psi} m_{B_c}^2 f_{B_c}}\bigg\{2{\cal S}\Theta (c(u,s_0)) \phi_{2;J/\psi}^\bot (u)-\frac{\lambda~ m_b ~m_{J/\psi} ~\tilde f_{J/\psi}}{u^2M^2}\hspace{1.2ex}\widetilde \Theta (c(u,s_0))
\nonumber\\
&\times& \int_0^udv \phi_{2;J/\psi}^\|(u) +m_{J/\psi}^2\bigg[{\cal Q}\,\Theta (c(u,s_0)) -  \frac\lambda{uM^2}\widetilde \Theta (c(u,s_0))\bigg] \psi_{3;J/\psi}^\| (u)  ~+~ m_b m_{J/\psi}\tilde f_{J/\psi}\hspace{0.8ex}\bigg[\frac{\lambda}{u^2M^2}
\nonumber\\
&\times& \widetilde\Theta(c(u,s_0)) \int_0^u dv \phi_{3;J/\psi}^\bot(u)+ {\cal Q}\Theta (c(u,s_0))\phi_{3;J/\psi}^\bot (u)\bigg] + m_{J/\psi}^2{\cal S}\bigg[\frac{{\cal R}}{2u^3M^4}\widetilde{\widetilde \Theta} (c(u,s_0)) \hspace{0.8ex}-\frac{3}{{2u^2M^2}}
\nonumber\\
&\times& \widetilde \Theta (c(u,s_0))\bigg] \phi_{4;J/\psi}^\bot (u) - \bigg[\frac{\lambda {\cal S}}{2u^3M^4}\widetilde{\widetilde \Theta} (c(u,s_0)) - m_{J/\psi}^2\frac{{\cal S} - 4\lambda }{u^2M^2}\bigg] \widetilde\Theta (c(u,s_0))I_L(u) -m_b^3m_{J/\psi}^3 \hspace{0.8ex}\tilde f_{J/\psi}
\nonumber\\
&\times& \frac{\lambda }{u^4M^6}~\widetilde{\widetilde \Theta} (c(u,s_0))\int_0^u dv~ \phi_{4;J/\psi}^\bot(u) +  m_b m_{J/\psi}^3{\tilde f_{J/\psi}}~ \bigg[\frac{\lambda }{{u^2M^4}}\widetilde{\widetilde \Theta} (c(u,s_0)) ~+ \frac{\cal Q}{u^2M^2}~\hspace{0.8ex}\widetilde \Theta (c(u,s_0))\,\bigg]
\nonumber\\
&\times& J_{J/\psi}(u) - m_{J/\psi}^2\bigg[ {\frac{3}{2}\Theta (c(u,s_0)) + \left( {\frac{{\cal R}}{u^2M^2} - \frac{\lambda}{{2uM^2{\cal Q}}}} \right) \widetilde \Theta (c(u,s_0))} \bigg]{H_3}(u)\bigg\},
\label{eq:HV0}
\\
\nonumber\\
{\cal H}_{1}^{B_c\to J/\psi}(q^2)
&=& \int_0^1 du e^{(m_{B_c}^2 - s)/M^2} \frac{m_b f_{J/\psi}^\bot \sqrt{2 q^2}} {2m_{B_c}^2 f_{B_c}} \bigg\{ \Theta (c(u,s_0))~\phi _{2;J/\psi}^\bot (u) + m_{J/\psi}^2~\bigg[\frac{{\cal R}}{u^3M^4}\widetilde{\widetilde\Theta} (c(u,s_0 )) ~ +\hspace{0.8ex} \frac{3}{u^2M^2}
\nonumber\\
&\times& \widetilde \Theta (c(u,s_0 ))\bigg]\phi_{4;J/\psi}^ \bot (u) - \frac{{m_{J/\psi}m_b{\tilde f_{J/\psi}}}}{{2u^2M^2}}\widetilde \Theta (c(u,s_0 ))\psi_{3;J/\psi}^\bot (u)\bigg\},
\label{eq:HV1}
\\
\nonumber\\
{\cal H}_{2}^{B_c\to J/\psi}(q^2)
&=& \int_0^1 ~du~ e^{(m_{B_c}^2 - s)/M^2} ~\frac{\sqrt {2q^2}~ m_b~ f_{J/\psi}^\bot}{2\sqrt \lambda m_{B_c}^2 f_{B_c}}~\bigg\{ {\cal P}~\Theta (c(u,s_0))~\phi_{2;J/\psi}^\bot(u) + 2m_{J/\psi}^2~\Theta(c(u,s_0))~\psi_{3;J/\psi}^\|(u)
\nonumber\\
&+ & m_{J/\psi}^2 {\cal P}\bigg[\frac{\cal R}{u^3M^4}\widetilde{\widetilde \Theta} (c(u,s_0)) + \frac{3}{u^2M^2}\widetilde \Theta (c(u,s_0))\bigg]\phi _{4;J/\psi}^ \bot (u)+ \frac{2m_{J/\psi}^2}{u^2 M^2} ~{\cal P}\widetilde \Theta (c(u,s_0))I_L(u) - m_{J/\psi}^2
\nonumber\\
&\times&\bigg[3\Theta (c(u,s_0)) + \frac{2{\cal R}}{u^2 M^2} \widetilde\Theta(c(u,s_0))\bigg]\!H_3(u) - 2m_bm_{J/\psi}\tilde f_{J/\psi} \Theta (c(u,s_0)) \phi_{3;J/\psi}^\bot (u)-\!2 m_b m_{J/\psi}^3 \tilde f_{J/\psi}
\nonumber\\
&\times& \frac{1}{u^2M^2} \widetilde\Theta (c(u,s_0)) J_{J/\psi}(u)\bigg\},
\label{eq:HV2}
\\
\nonumber\\
{\cal H}_t^{B_c\to J/\psi}(q^2)
&=&  \int_0^1 du ~ e^{(m_{B_c}^2 - s)/M^2} \frac{m_b~m_{J/\psi}~f_{J/\psi}^\bot}{2m_{J/\psi} m_{B_c}^2 f_{B_c}} ~\bigg\{ u~m_{J/\psi} \Theta (c(u,s_0))\phi_{2;J/\psi}^\bot(u) - m_b \tilde f_{J/\psi}~\frac{ \cal Q}{u^2 M^2}~\widetilde\Theta(c(u,s_0))
\nonumber\\
&\times&\int_0^u dv \phi_{2;J/\psi}^\|(u) - (\tilde m_q \tilde f_{J/\psi}- m_{J/\psi} )~\bigg[\Theta (c(u,s_0)) + \frac{{u{\cal Q} + 2{q^2}}}{u^2 M^2}\widetilde \Theta (c(u,s_0))\bigg]\psi _{3;J/\psi}^\| (u) - m_b \tilde f_{J/\psi}
\nonumber\\
&\times& \Theta (c(u,s_0))~\phi_{3;J/\psi}^\bot(u) + m_b \tilde f_{J/\psi} ~\frac{\cal Q}{u^2 M^2}~\widetilde\Theta(c(u,s_0))\int_0^u dv~ \phi_{3;J/\psi}^\bot(u) + m_{J/\psi}^3\bigg[\frac{{\cal R}}{u^2 M^4}\widetilde {\widetilde \Theta }(c(u,s_0))
\nonumber
\\
&+&\frac3{uM^2}\widetilde \Theta (c(u,s_0))\bigg]\phi _{4;J/\psi}^ \bot (u) ~+~ m_b^3m_{J/\psi}^2 \tilde f_{J/\psi} \frac{{\cal Q}}{u^4M^6}\widetilde {\widetilde {\widetilde \Theta }}(c(u,s_0))\int_0^u dv \,\phi_{4;J/\psi}^\bot(u) -  m_{J/\psi} \bigg[\frac{{\cal P}}{{2u{M^2}}}
\nonumber\\
&\times& \widetilde \Theta (c(u,s_0)) ~+~ \frac{3}{2}~\Theta (c(u,s_0))\bigg]~H_3(u) ~-~ m_{J/\psi}\bigg[~\frac{{9{\cal Q} \,-\, 2um_{J/\psi}^2 \,+\, 15{q^2}}}{u^2 M^2}~\widetilde \Theta (c(u,s_0)) ~+~ \frac{{\cal T}}{u^3 M^4}
\nonumber\\
&\times& \widetilde {\widetilde \Theta }(c(u,s_0))\bigg]{I_L}(u) + \frac{{m_b \tilde f_{J/\psi} }}{2}\bigg[\frac{{2m_{J/\psi}^2}}{u^2 M^2}\widetilde \Theta (c(u,s_0)) + \frac{{\cal S}}{u^3 M^4}\widetilde {\widetilde \Theta }(c(u,s_0))\bigg]J_{J/\psi}(u)\bigg\},\label{eq:HVt}
\end{eqnarray}
\end{widetext}
where $\tilde f_{J/\psi} = f_{J/\psi}^\| / f_{J/\psi}^\bot$, ${\cal P} = m_{B_c}^2 + \xi m_{J/\psi}^2 - q^2$, ${\cal Q} = m_{B_c}^2 - m_{J/\psi}^2 - q^2$, ${\cal R} = u m_{B_c}^2 - u \bar u m_{J/\psi}^2 + \bar u{q^2}$, ${\cal S} = 2m_{J/\psi}^2(um_{B_c}^2 - um_{J/\psi}^2 + (1 - \bar u){q^2})$, ${\cal T} = 2m_{B_c}^2[ - u\xi m_{J/\psi}^2 + {q^2}(1 + u + u\bar u)] + u\xi (m_{B_c}^4 + m_{J/\psi}^4) - 2{q^2}(1 + u)\bar um_{J/\psi}^2 - {q^4}(2 + u)$ and $s=[m_b^2-\bar u(q^2-u m_{J/\psi}^2)]/u$ with $\bar u=1-u$, $\xi = 2u-1$. The $\Theta(c(u,s_0))$, $\widetilde\Theta [c(u,s_0)]$, $\widetilde{\widetilde\Theta}(c(u,s_0)]$ and $\widetilde{\widetilde{\widetilde\Theta}}(c(u,s_0)]$ are the step function. The simplified distribution functions $H_3(u)$, $I_L(u)$ and $J_{J/\psi}(u)$ can be written as
\begin{align}
&H_3(u) = \int_0^u dv \Big[\psi_{4;J/\psi}^\bot(v)-\phi_{2;J/\psi}^\bot(v)\Big], \nonumber\\
&I_L(u) = \int_0^u dv \int_0^v dw \Big[\phi_{3;J/\psi}^\|(w) -\frac{1}{2} \phi_{2;J/\psi}^\bot(w)
\nonumber\\
&\hspace{5.6ex} -\frac{1}{2} \psi_{4;J/\psi}^\bot(w)\Big], \nonumber \\
&J_{J/\psi}(u) = \int_0^u {dv} \int_0^v {dw} \Big[ \psi _{4;J/\psi}^\| (w) + \phi_{2;J/\psi}^\| (w)
\nonumber\\
&\hspace{8ex} - 2\phi _{3;J/\psi}^ \bot (w) \Big].
\end{align}
A more detailed calculation can be found in our previous work~\cite{Cheng:2018ouz,Fu:2020vqd}. Since we used a usual current in the calculation, the HFFs will include all the Twist-2,-3,-4 light-cone distribution amplitudes (LCDAs), see subsection~\ref{subsection:LCDAandHFF}, which will maintain the accuracy of LCSR to a large extent.

\section{Numerical results and discussion} \label{Section:III}

While performing the numerical calculations, we take the mass of ${B_c}$- and $J/\psi$-meson are $m_{B_c}=6.2749~\rm{GeV}$ and $m_{J/\psi}=3.097~\rm{GeV}$~\cite{Tanabashi:2018oca}, ${B_c}$-meson decay constant $f_{B_c}=0.498\pm{0.014}~\rm{GeV}$~\cite{Zhong:2014fma}, $J/\psi$-meson decay constant  $f_{J/\psi}^\perp = 0.410\pm{0.014} ~\rm{GeV}$ and $f_{J/\psi}^\|=0.416\pm{0.005}~\rm{GeV}$~\cite{Becirevic:2013bsa,Zeng:2021hwt}. The CKM-matrix elements will set its central values, i.e., $|V_{cb}| = 0.0405$, $|V_{ud}| = 0.974$ and $|V_{us}| = 0.225$~\cite{Patrignani:2016xqp}. For the factorization scale $\mu$, we will fix it as the typical momentum transfer of $B_c \to J/\psi$~\cite{Ball:2004rg}, i.e., $\mu\simeq(m_{B_c}^2-m_b^2)^{1/2}\sim4.68~\rm{GeV}$.

\subsection{$J/\psi$-meson LCDAs and $B_c \to J/\psi$ HFFs}\label{subsection:LCDAandHFF}

An important part in the LCSR $B_c \to J/\psi$ HFFs is the $J/\psi$-meson LCDAs. For the leading twist DAs $\phi_{2,J/\psi}^\lambda(x,\mu)$, we will take the Wu-Huang (WH) model to carry out our analysis. Its definition is as follows~\cite{Zeng:2021hwt}:
\begin{align}
\phi_{2;J/\psi}^\lambda(x,\mu)&=\frac{\sqrt{3}A_{J/\psi}^\lambda \hat m_c\beta_{J/\psi}^\lambda\sqrt{x\bar{x}}}{2\pi^{3/2}f_{J/\psi}^\lambda}\bigg\{{\rm Erf}\bigg[\sqrt\frac{\hat m_c^2+\mu^2}{8(\beta_{J/\psi}^{\lambda})^2x\bar{x}}\bigg]
\nonumber\\
&-{\rm Erf}\bigg[\sqrt\frac{\hat m_c^2}{8(\beta_{J/\psi}^{\lambda})^2x\bar{x}}\bigg]\bigg\},
\label{eq:DABHL}
\end{align}
where $\lambda=(\bot,\|)$, $\bar x=(1-x)$, the $c$-quark mass is taken as $\hat m_c =1.5~{\rm GeV}$, and the error function ${\rm Erf}(x)=2\int_0^x e^{-t^2}dt /\sqrt{\pi}$. The remaining model parameters $A_{J/\psi}^\lambda$ and $\beta_{J/\psi}^\lambda$ can be fixed by employing the normalization condition, i.e., $
\int \phi_{2;J/\psi}^{\lambda}(x,\mu)dx = 1$, and the second Gegenbauer moment $a_{2;J/\psi}^\|(\mu_0)=-0.379\pm{0.020}$ and $a_{2;J/\psi}^\bot (\mu_0)=-0.373_{-0.015}^{+0.026}$~\cite{Braguta:2007fh} that are related to the leading-twist DAs, i.e.,
\begin{align}
a_{2;J/\psi}^\lambda (\mu) = \dfrac{\displaystyle\int_0^1 dx \phi_{2;J/\psi}^{\lambda} (x,\mu) C_2^{3/2}(2x-1)}{\displaystyle\int_0^1 6x \bar x [C_2^{3/2}(2x-1)]^2}.
\label{eq:anphi}
\end{align}

The final model parameters are listed in Table.~\ref{tab:Model_p}. Note that the model parameters at arbitrary scale can be obtained by running the scale
dependence of the Gegenbauer moments~\cite{Ball:2006nr}.

\begin{table}[t]
\caption{The determined model parameters of leading twist-2 LCDA with $\mu_0 = 1.2~{\rm GeV}$.}
\begin{center}\label{tab:Model_p}
\renewcommand{\arraystretch}{1.4}
\begin{tabular}{ cccccc }
\hline
~$a_{2;J/\psi}^\| (\mu_0)$ ~& ~$A_{J/\psi}^\|$ ~&~ $\beta_{J/\psi}^\|$ ~& $a_{2;J/\psi}^\bot (\mu_0)$ &~ $A_{J/\psi}^\bot$ ~&~ $\beta_{J/\psi}^\bot$~\\
\hline
 $-0.379$ & $3818$ & $0.536$ & $-0.373$& $3010$ & $0.548$\\
 $-0.359$ & $1841$ & $0.579$ & $-0.353$& $1510$ & $0.592$\\
 $-0.400$ & $9341$ & $0.493$ & $-0.394$& $6988$ & $0.506$\\
\hline
\end{tabular}
\end{center}
\end{table}

For the $J/\psi$-meson twist-3 LCDAs, we will relate it to leading-twist DAs by employing the Wandzura-Wilczek approximation~\cite{Wandzura:1977qf, Ball:1997rj}. The specific relationships are as follows:
\begin{align}
\psi_{3;J/\psi}^\bot(u)&=2~\bigg[~\bar u \int_0^u dv \frac {\phi_{2;J/\psi}^\|(v)}{\bar v}+ u \int_u^1 dv \frac {\phi_{2;J/\psi}^\|(v)}{v}\hspace{0.2ex}\bigg],\nonumber\\
\phi_{3;J/\psi}^\bot(u)&=\frac12~\bigg[~\int_0^u ~dv~ \frac {\phi_{2;J/\psi}^\|(v)}{\bar v}\,+\int_u^1~ dv \frac {\phi_{2;J/\psi}^\|(v)}{v}\bigg],\nonumber\\
\psi_{3;J/\psi}^\|(u)&=2\,\bigg[\,\bar u \,\int_0^u dv \frac {\phi_{2;J/\psi}^\bot(v)}{\bar v}+ u \,\int_u^1 dv \frac {\phi_{2;J/\psi}^\bot(v)}{v}\hspace{0.2ex}\bigg],\nonumber\\
\phi_{3;J/\psi}^\|(u)&=(1-2\bar u)\!\bigg[\!\int_0^u\!\!dv \frac {\phi_{2;J/\psi}^\bot(v)}{\bar v}-\int_u^1\!\!dv \frac {\phi_{2;J/\psi}^\bot(v)}{v}\bigg],\nonumber\\
\end{align}
where $\bar{u} =(1-u)$ and $\bar{v} =(1-v)$. For the $J/\psi$-meson twist-4 LCDAs, as its contributions are usually small in comparison to that of the twist-2,-3, we shall ignore charm-quark mass effect of the twist-4 LCDAs~\cite{Ball:1998kk} to do our analysis.
\begin{figure*}[htb]
\centering
\includegraphics[width=0.25\textwidth]{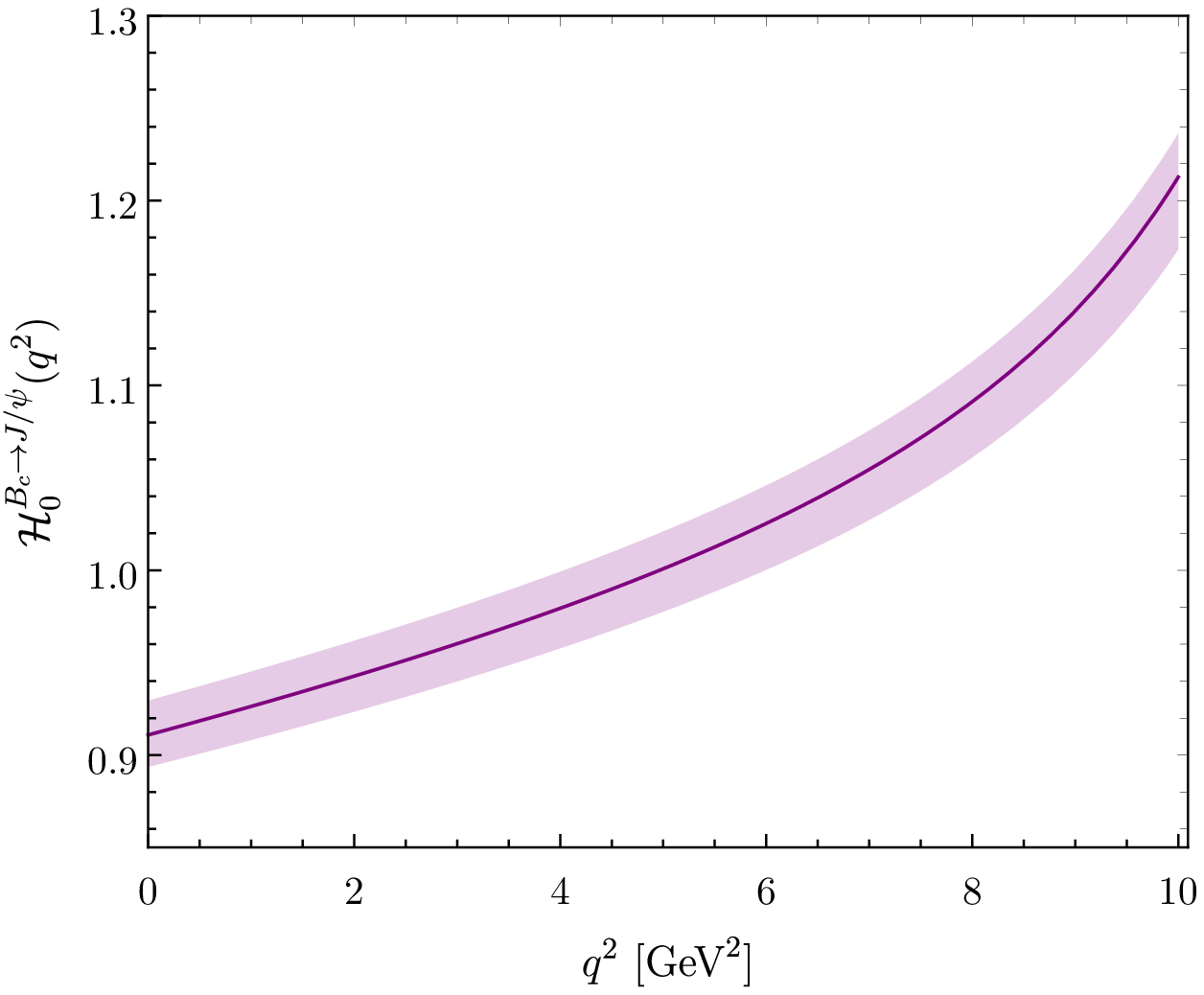}\includegraphics[width=0.25\textwidth]{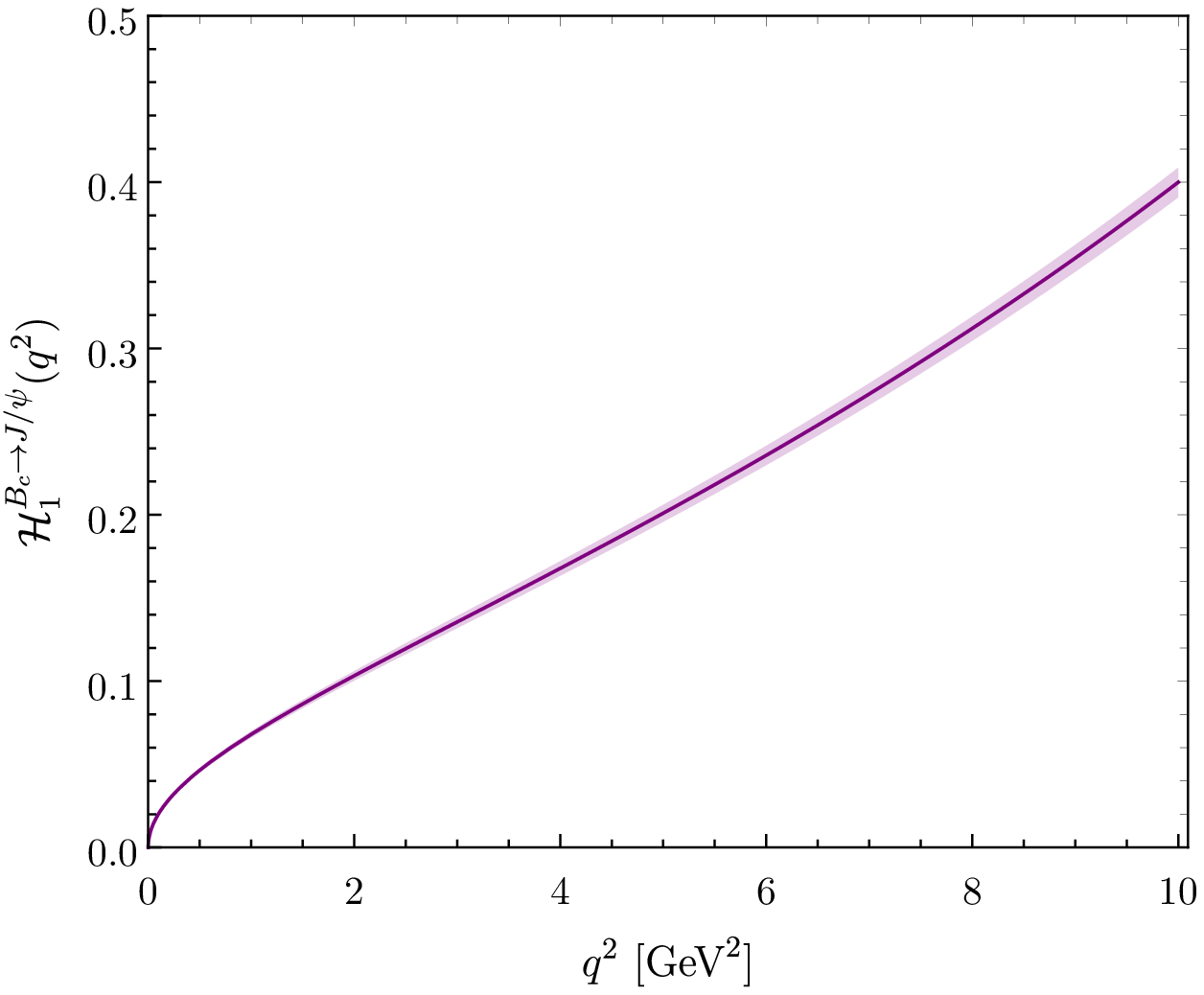}\includegraphics[width=0.245\textwidth]{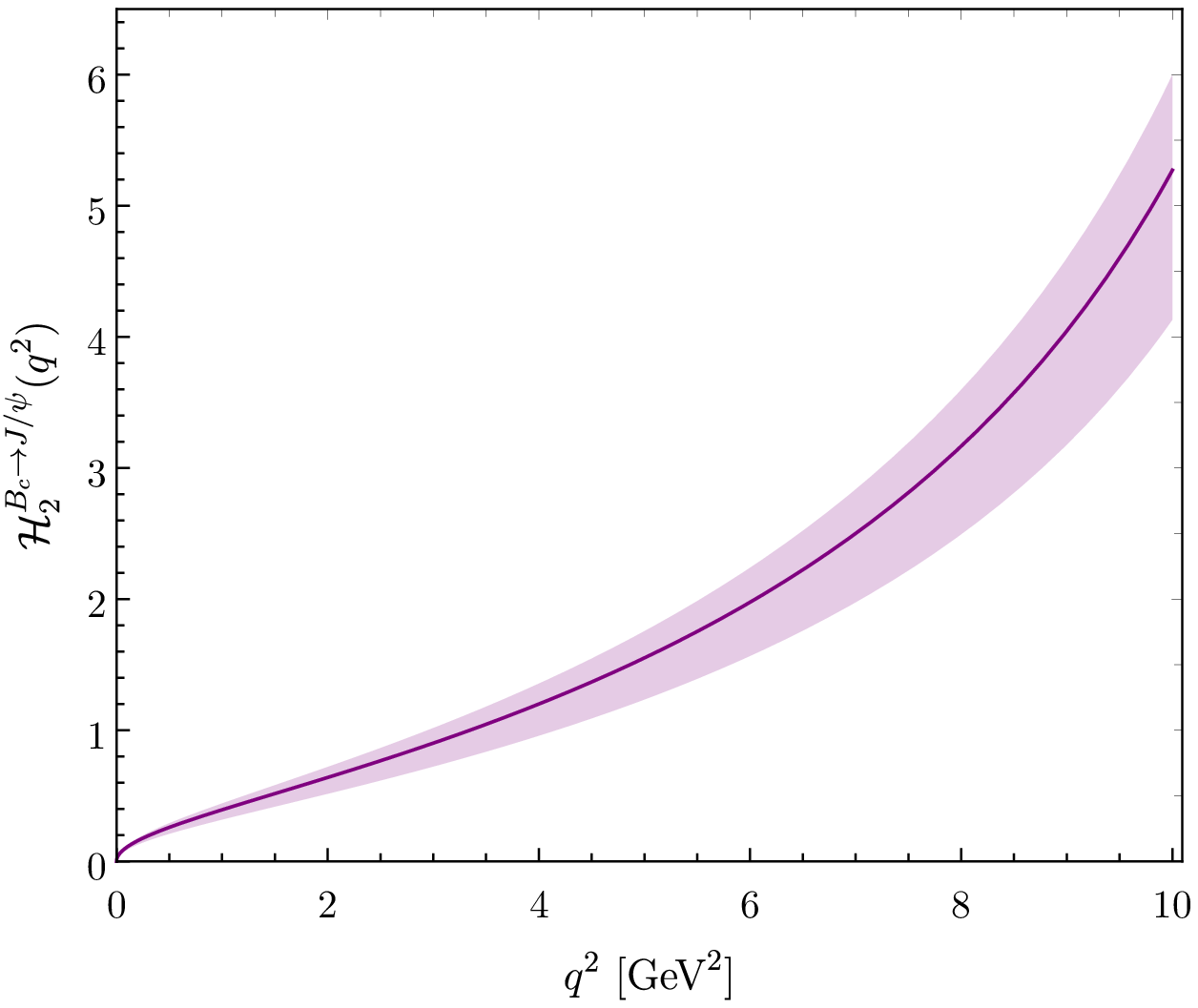}\includegraphics[width=0.25\textwidth]{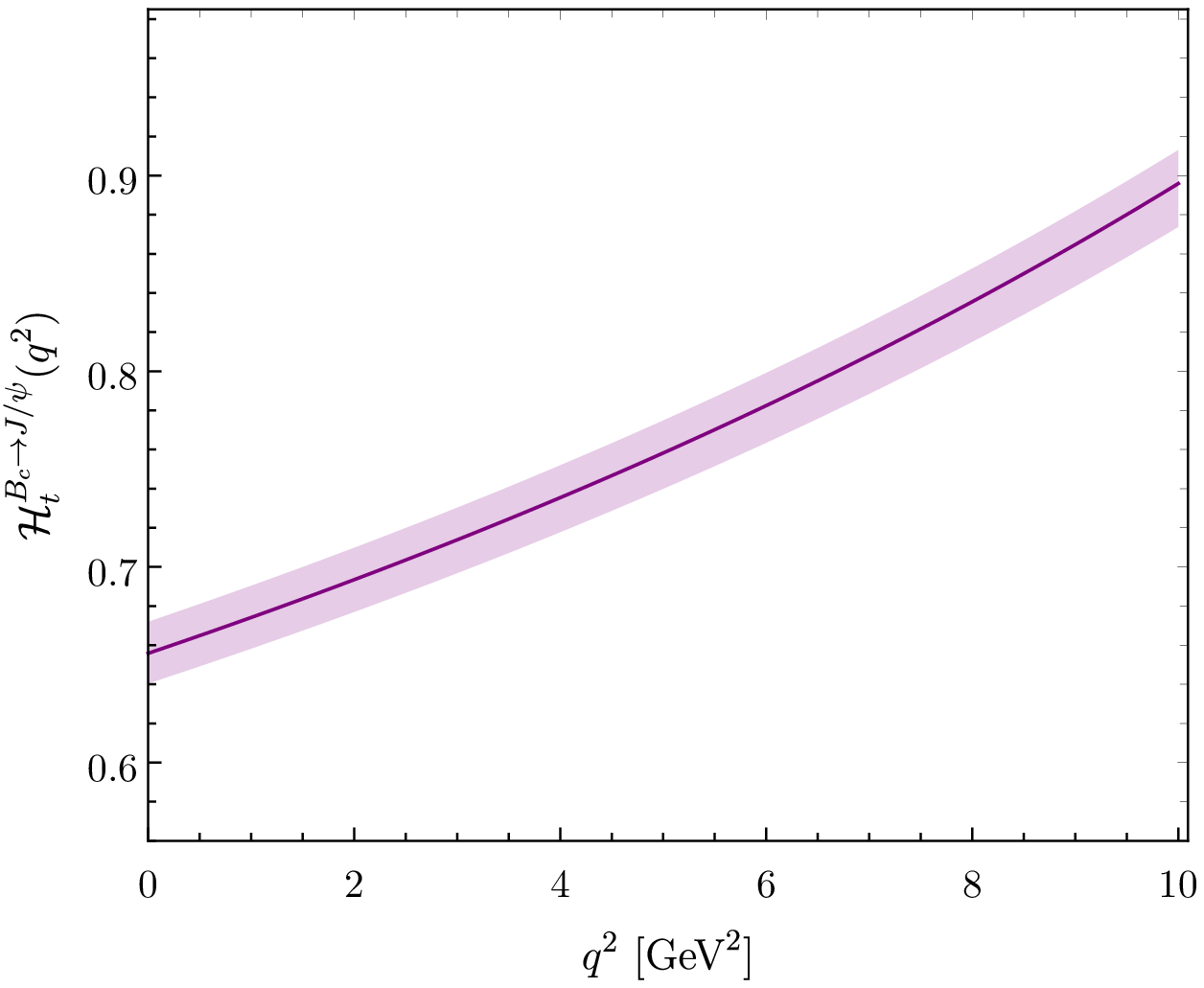}
\caption{The extrapolated HFFs ${\cal H}_\sigma^{B_c\to J/\psi}(q^2)$ as a function of $q^2$, in which the solid lines represent the center values and
the shaded bands stand for the uncertainties that are the squared of all mentioned error sources. }
\label{Fig:HFF012t}
\end{figure*}

As for the continuum threshold $s_0$ of the $B_c\to J/\psi$ HFFs ${{\cal H}_\sigma^{B_c \to J/\psi}}(q^2)$, it can usually be set near the squared mass of $B_c$-meson's first exciting state or the value between $B_c$-meson ground state and higher mass contributions. We will fix it as $s_{{\cal H}_\sigma}=45.0(0.5)~\rm{GeV}^2$. In contrast, to obtain the Borel windows of the $B_c\to J/\psi$ HFFs, we require the continuum contribution should be less than $65\%$ of the total LCSR. Thus, the final Borel windows $M_{{\cal H}_{j}}^2 (\rm{GeV}^2)$ are $M^2_{{\cal H}_{0}}=71.0(1.0)$, $M^2_{{\cal H}_{1}}=72.0(1.0)$, $M^2_{{\cal H}_{2}}=8.5(1.0)$, and $M^2_{{\cal H}_{t}}=180.0(1.0)$.

Compared with the TFFs decomposition method, for a single TFFs composed of helicity amplitude, the value of $s_0$ and $M^2$ of the two methods have no affect to the helicity amplitude. However, for a mixture of multiple TFFs composed of helicity amplitude, the value of $s_0$ and $M^2$ of different TFFs is usually different, therefore, there will be a difference in the helicity amplitude obtained by the two methods, which will affect the following theoretical prediction.

Since the reliable region of the LCSR is in the lower and intermediate $q^2$-region, we can take it as $0\leq q^2\leq q^2_{\rm LCSR, max}\approx 5~\rm{GeV}^2$ for the $B_c\to J/\psi$ decay, and the allowable physical range of the momentum transfer is $0\leq q^2\leq q^2_{B_c\to J/\psi,{\rm max}}$ with $q^2\leq q^2_{B_c\to J/\psi,{\rm max}}  = (m_{B_c}-m_{J/\psi})^2 \simeq 10.10~\rm{GeV}^2$. Therefore, the HFFs obtained by LCSR need to be extended into full allowable physical ranges, so that  the  $B_c$-meson decays can be studied further. Here, we will adopt the SSE method~\cite{Bourrely:2008za,Bharucha:2010im} to perform the extrapolation due to the analyticity and unitarity consideration. The extrapolated formulas of the $B_c\to J/\psi$ HFFs ${\cal H}_\sigma^{B_c\to J/\psi}(q^2)$ can be expressed as:

\begin{table}[t]
\caption{Relations between the mass of low-lying $B_c$ resonances and the parametrization of HFFs ${\cal H}_\sigma^{\rm fit}$.}
\begin{center}
\renewcommand{\arraystretch}{1.4}
\begin{tabular}{llll}
\hline
Transition~~~~~~~& $J^P$~~~~~~~~~~~~~~~ & $m_{R,i}$ (GeV) ~~~ & ${\cal H}_{\sigma}^{\rm fit}(q^2)$  \\
\hline
            & $0^-$  & 6.28  & ${\cal H}_{t}^{\rm fit}(q^2)$  \\
$b\to c$& $1^-$  & 6.34  & ${\cal H}_{1}^{\rm fit}(q^2)$  \\
            & $1^+$ & 6.75  & ${\cal H}_{0,2}^{\rm fit}(q^2)$\\
\hline
\end{tabular}
\label{tab:reson}
\end{center}
\end{table}

\begin{eqnarray}
{\cal H}_{0}^{\rm fit}(t) &=&\frac{1}{B_{0}(t) \sqrt{z(t,t_-)} \phi_T^{V-A}(t)} \sum_{k=0,1,2} a_k^{0} z^k , \label{Eq:extraHHFs1} \\
{\cal H}_{1}^{\rm fit}(t) &=&\frac{\sqrt{-z(t,0)}}{B_{1}(t) \phi_T^{V-A}(t)} \sum_{k=0,1,2} a_k^{1} z^k ,   \label{Eq:extraHHFs2} \\
{\cal H}_{2}^{\rm fit}(t) &=&\frac{\sqrt{-z(t,0)}}{B_{2}(t) \sqrt{z(t,t_-)} \phi_T^{V-A}(t)} \sum_{k=0,1,2} a_k^{2} z^k ,\label{Eq:extraHHFs3} \\
{\cal H}_{t}^{\rm fit}(t) &=&\frac{1}{B_{t}(t)\phi_L^{V-A}(t)} \sum_{k=0,1,2} a_k^{t} z^k ,\label{Eq:extraHHFs4}
\end{eqnarray}
where $B_i(t)=1- q^2/m_i^2$ with $i=(0,1,2,t)$. The mass $m_{0,1,2,t}$ are in Table.~\ref{tab:reson}, $\phi_{T,L}^{V-A}(t)=1$, $\sqrt{-z(t,0)}=\sqrt{q^2}/m_{B_c}$, $\sqrt{z(t,t_-)}=\sqrt{\lambda(q^2)}/m_{B_c}^2$, and $z(t)=
({\sqrt{t_+ - t}-\sqrt{t_+ - t_0}})/({\sqrt{t_+ - t}+\sqrt{t_+ - t_0}})$ with $t_\pm=(m_{B_c} \pm m_{J/\psi})^2$ and $t_0=t_+(1-\sqrt{1-t_-/t_+})$.

\begin{table}[t]
\caption{The fitted results of the parameters $a^{\sigma}_{k}$ and $\Delta$ for the HFFs ${\cal H}_{\sigma}^{B_c \to J/\psi}$, where we take all input parameters as their central values.}
\begin{center}\label{tab:deltaanda}
\renewcommand{\arraystretch}{1.4}
\begin{tabular}{ccccc}
\hline
&${\cal H}_0^{B_c\to J/\psi}(q^2)$ & ${\cal H}_1^{B_c\to J/\psi}(q^2)$ & ${\cal H}_{2}^{B_c\to J/\psi}(q^2)$ &${\cal H}_t^{B_c\to J/\psi}(q^2)$\\
\hline
$a_0^{\sigma}$    & $0.471$   & $0.507$  & $2.819$    & $0.662$\\
$a_1^{\sigma}$    & $13.409$ & $-6.746$ & $-53.780$ & $-0.428$\\
$\Delta$                & 0.927      & $0.917$   & $0.991$    & $0.992$\\
\hline
\end{tabular}
\end{center}
\end{table}

\begin{table*}[th]
\caption{Decay width of the two body decays $B_c^+ \to J/\psi  +(P, V)$ in units $a_1^2 \times 10^{-15}~\rm{GeV}$.}
\begin{center}
\renewcommand{\arraystretch}{1.4}
\begin{tabular}{lcccc}
\hline
~~~~~~~~~~~~~~~~~~~~~&~~~~~~~$B_c^+ \to J/\psi \pi^+$~~~~~~~&~~~~~~$B_c^+ \to J/\psi \rho^+$~~~~~~~&~~~~~~~$B_c^+ \to J/\psi K^+$ ~~~~~~~~~&  $B_c^+ \to J/\psi K^{\ast +}$\\
\hline
This Work &$1.642^{+0.082}_{-0.076}$    & $9.251^{+0.560}_{-0.604}$        &  $0.123^{+0.006}_{-0.006}$ 	 & $0.524^{+0.035}_{-0.039}$\\
RM~\cite{AbdElHady:1999xh}&$ 1.22 $  	       & $ 3.48 $  	                         &  $ 0.09 $  	                     &  $ 0.20 $ \\
RQM~\cite{Ebert:2003cn}	   &$ 0.67 $  	       & $ 1.8 $  	                         &  $ 0.05 $ 	                     &  $ 0.11$ \\
CCQM~\cite{Issadykov:2018myx}   &$ 1.22\pm 0.24 $  	           & $ 2.03\pm 0.41 $  	                 &  $ 0.09\pm 0.02  $ 	             &  $ 0.13\pm0.03$ \\
BS RQM~\cite{Chang:2014jca}   &$ 1.24\pm0.11 $    & $ 3.59^{+0.64}_{-0.58} $  	         &  $ 0.095\pm0.008 $ 	             &  $ 0.226\pm0.03 $ \\
\hline
\end{tabular}\label{tab:DwKpi}
\end{center}
\end{table*}
\begin{table*}[t]
\caption{Branching fractions (in unit of $\%$) of the two body decays $B_c^+ \to J/\psi +(P, V)$ decays obtained with $a_1 =1.038$.}
\begin{center}
\renewcommand{\arraystretch}{1.4}
\begin{tabular}{ccccc}
\hline
~~~~~~~~~~~~~~~~~~~~~&~~~~~~~$B_c^+ \to J/\psi \pi^+$~~~~~~~&~~~~~~$B_c^+ \to J/\psi \rho^+$~~~~~~~&~~~~~~~$B_c^+ \to J/\psi K^+$ ~~~~~~~~~&  $B_c^+ \to J/\psi K^{\ast +}$
\\
\hline
This Work            &$0.136^{+0.002}_{-0.002}$    & $0.768^{+0.029}_{-0.033}$   &  $0.010^{+0.000}_{-0.000}$ 	 & $0.043^{+0.001}_{-0.001}$\\
CCQM~\cite{Issadykov:2018myx}                &$0.101\pm0.02$                  &  $0.334\pm 0.067$  	           &  $0.008\pm 0.002$ 	             &  $0.019\pm 0.004$\\
\hline
\end{tabular}\label{tab:BrBctoKpi}
\end{center}
\end{table*}
To fix the parameters $a_k^{\sigma}$, we will take the ``quality'' of fit ($\Delta$) to be less than 1\%, i.e.,
\begin{equation}
\Delta=\frac{\sum_t|{\cal H}_{\sigma}^{B_c\to J/\psi}(t)-{\cal H}_{\sigma}^{\rm fit}(t)|} {\sum_t|{\cal H}_{\sigma}^{B_c\to J/\psi}(t)|}\times 100\%,
\end{equation}
where $t\in[0,0.5,\cdots,4.5,5.0]~{\rm GeV}^2$. The determined parameters $a_{k}^{\sigma}$ are listed in Table~\ref{tab:deltaanda}, in which all the input parameters are set to be their central values. With the extrapolated Eq.~(\ref{Eq:extraHHFs1})-(\ref{Eq:extraHHFs4}) and the fitted parameters $a^{\sigma}_{k}$, we show the extrapolated HFFs ${\cal H}_\sigma^{B_c\to J/\psi}(q^2)$ in whole $q^2$-regions in Fig.~\ref{Fig:HFF012t}, where the shaded band stands for the squared average of all the mentioned uncertainties. Same as our previous study on pseudoscalar meson decay to vector meson, such as $B \to \rho$~\cite{Cheng:2018ouz} and $D \to V$~\cite{Fu:2020vqd}, the transverse part of the final meson does not contribute to HFFs at the large recoil point $q^2=0$. Specifically, we have ${\cal H}_{0}^{B_c\to J/\psi}(0)=0.915_{-0.018}^{+0.022}$, ${\cal H}_{t}^{B_c\to J/\psi}(0)=0.644_{-0.015}^{+0.018}$, and ${\cal H}_{1,2}^{B_c\to J/\psi}(0)=0$, where the errors are from the $J/\psi$-meson decay constant $f_{J/\psi}$, the $J/\psi$-meson LCDAs, the Borel parameter $M^2$, and the continuum threshold $s_0$. All the HFFs ${\cal H}_\sigma^{B_c\to J/\psi}(q^2)$ are monotonically increase with the increment of $q^2$.

\subsection{The two body decays $B_c^+ \to J/\psi +(P, V)$ }

\begin{table*}

\caption{The branching ratios ${\cal R}_{\pi^+/ \mu^+\nu_\mu}$, ${\cal R}_{K^+ /\pi^+}$ and ${\cal R}_{J/\psi}$ with the errors are squared average from various input parameters.}
\begin{center}
\begin{tabular}{ llll}
\hline
~~~~~~~~~~~~~~~~~~~~~~~~~~~~~~~~~~~~~~~~~& $\mathcal{R_{\pi^+/ \mu^+ \nu_\mu}}$~~~~~~~~~~~~~~~~~~~~~~~~~~~~ 	& ${\cal R}_{K^+ /\pi^+}$~~~~~~~~~~~~~~~~~~~~~~~~~~~    & ${\cal R}_{J/\psi}$\\
\hline
This work			     	                   &$0.048^{+ 0.009}_{-0.012}$ &$0.075^{+ 0.005}_{-0.005}$        &$0.199^{+ 0.060}_{-0.077}$ \\
LHCb~\cite{Aaij:2014jxa}	            & $0.047 \pm0.005$	                &                                  &                           \\
LHCb~\cite{Aaij:2013vcx}            &                                            & $0.069 \pm 0.019$                &                           \\
LHCb~\cite{Aaij:2016tcz} 	        &                                            & $ 0.079 \pm0.008$               &                           \\
LHCb~\cite{Aaij:2017tyk}            &                                            &                                      &$ 0.71\pm0.25$             \\
RM~\cite{AbdElHady:1999xh}          &$ 0.0525 $			                    &$0.074$                           &                           \\
LFQM~\cite{Ke:2013yka} 	           &$ 0.058 $			    	            &$0.075$                           &                           \\
CCQM~\cite{Issadykov:2018myx}&$0.061 \pm 0.012$			        &$0.076\pm0.015$              &$0.24\pm0.05$              \\
BS RQM~\cite{Chang:2014jca}	   &$0.064^{+0.007} _{-0.008}$ &$0.072^{+0.019} _{-0.008} $       &                           \\
RQM~\cite{Ebert:2003cn}           &$ 0.050 $			    	               &$0.077$                           &                           \\
pQCD~\cite{Wen-Fei:2013uea}	   &                                            &                                  &$0.29$                     \\
LCSR~\cite{Leljak:2019eyw}	        &                                            &                 & $ 0.23\pm0.01$                          \\
QCD-SR~\cite{Azizi:2019aaf}    &                                            &                                  &$0.25\pm0.01$                     \\
CLFQM~\cite{Huang:2018nnq}      & 			                            &                      &$0.248\pm0.006$                          \\
pQCD~\cite{Shen:2014msa}	       &                                            &                                  &$0.26$                     \\
NRQCD~\cite{Qiao:2012hp}        &          			                       &$ 0.075 $                         &                           \\
\hline
\end{tabular}\label{tab:Rall}
\end{center}
\end{table*}

\begin{figure}[h]
\begin{center}
\includegraphics[width=0.4\textwidth]{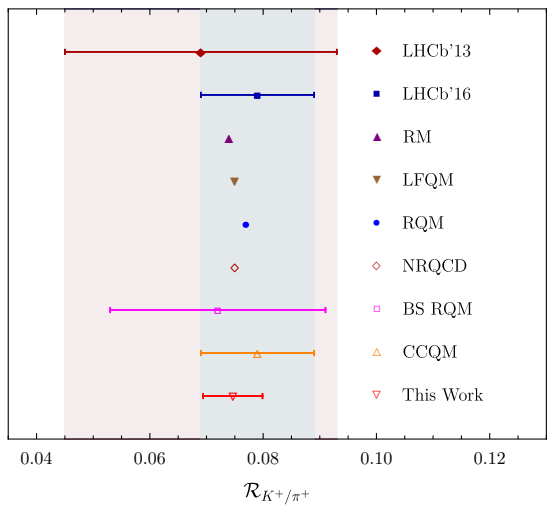}
\end{center}
\caption{LCSR prediction for the branching ratio ${\cal R}_{K^+/\pi^+}$.}
\label{R:Kpi}
\end{figure}

\begin{table}[t]

\caption{Branching fractions (in unit of $\%$) of the $B_c^+$-meson semileptonic decays within uncertainties. Other theoretical predictions are also listed here to make a comparison.}\label{tab:BFsemi}
\begin{center}
\begin{tabular}{lll}
\hline
~~~~~~~~~~~~~~~~~~~~~~~~~&$B_c^+ \to J/\psi \mu^+\nu_\mu$~~~~~~ &  $B_c^+ \to J/\psi \tau^+\nu_\tau$
\\
\hline
This Work            &$2.802^{+0.526}_{-0.675}$    & $0.559^{+0.131}_{-0.170}$  \\
RM~\cite{AbdElHady:1999xh}                  &$2.01$                         & \\
CCQM~\cite{Issadykov:2018myx}               &$1.67\pm0.33$                  & $0.40\pm 0.08$  	           \\
BSM~\cite{Chang:1992pt}                 &$2.33$                         &   	           \\
BS RQM~\cite{Chang:2014jca}                 &$1.73$                         &   	           \\
RQM~\cite{Ebert:2003cn}                     &$1.67\pm0.33$                  &$1.23$  	       \\
pQCD~\cite{Wen-Fei:2013uea}                 &$1.01$                         &$0.29$    	       \\
LCSR~\cite{Leljak:2019eyw}                    &$2.24^{+0.57}_{-0.49}$         &$0.53^{+0.16}_{-0.14}$  \\
QCD-SR~\cite{Azizi:2019aaf}                 &$1.93^{+0.50}_{-0.60}$         &$0.49^{+0.10}_{-0.14}$    	       \\
\hline
\end{tabular}
\end{center}
\end{table}

After setting all the input parameters and getting the extrapolated $B_c \to J/\psi$ HFFs, we can now analyze the $B_c$-meson decays numerically. For the $B_c$-meson two-body decay $B_c^+ \to J/\psi +(P, V)$, its decay width can be calculated by employing the Eq.~\eqref{Eq:DW}. We list our results in Table.~\ref{tab:DwKpi}. As a comparison, RM~\cite{AbdElHady:1999xh}, CCQM~\cite{Issadykov:2018myx}, BS RQM~\cite{Chang:2014jca} and RQM~\cite{Ebert:2003cn} predictions are also shown. After further taking into account the $B_c$-meson lifetime $\tau_{B_c}^{\rm exp}=(0.507\pm0.009)~{\rm ps}$ from Particle Data Group~\cite{Zyla:2020zbs}, we can get the corresponding branching fractions that are listed in Table.~\ref{tab:BrBctoKpi}. In order to compare with the experiments, we calculate the branching ratio ${\cal R}_{K^+/\pi^+}$, which is given by
\begin{equation}
\label{eq:LHCb-2}
{\cal R}_{K^+ / \pi^+}=
\frac { {\cal B}(B_c^+ \to J/\psi K^+)}
      {{\cal B}(B_c^+ \to J/\psi \pi^+)}.
\end{equation}

The numerical results are listed in Table.~\ref{tab:Rall}. Although our results related to the two-body decay width in Table.~\ref{tab:DwKpi} and branching fractions in Table.~\ref{tab:BrBctoKpi} are larger than that of other theories, our results ${\cal R}_{K^+ / \pi^+}$ are consistent with the latest LHCb experiment results within the $1\sigma$-errors. For a more intuitive comparison, we show ${\cal R}_{K^+ / \pi^+}$ in Fig.~\ref{R:Kpi}.

\subsection{The semi-leptonic decays $B_c^+ \to J/\psi \ell^+ \nu_\ell$}

\begin{figure*}[t]
\begin{center}
\includegraphics[width=0.4\textwidth]{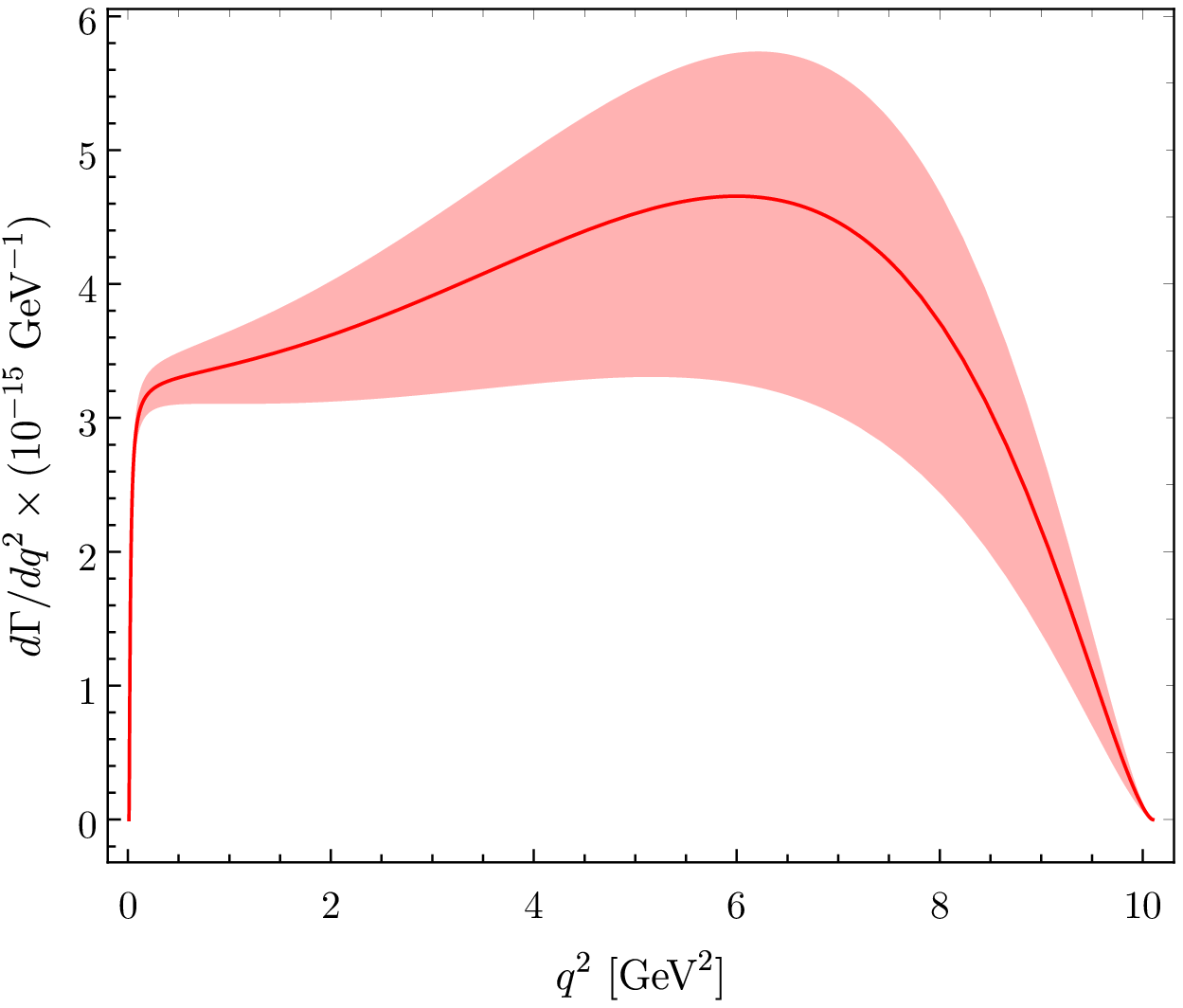}
\includegraphics[width=0.41\textwidth]{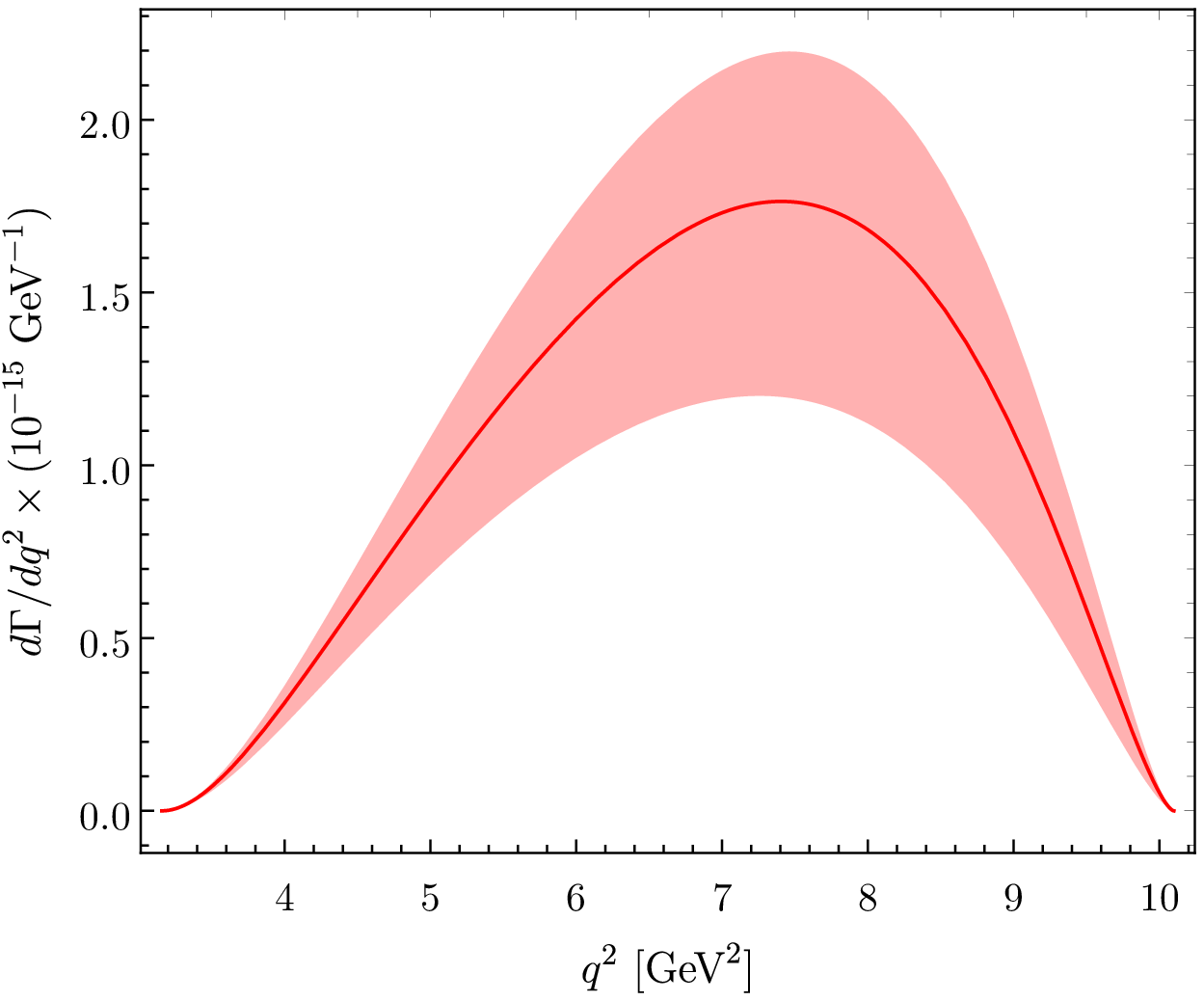}
\end{center}
\caption{The differential decay width of the semileptonic $B_c^+ \to J/\psi  \ell^+ \nu_\ell$ decays with $\ell=\mu$ for left and $\ell=\tau$ for right, where the solid and shade bands correspond to their central values and the uncertainties respectively.}
\label{DW:mutaup}
\end{figure*}

\begin{figure*}[th]
\begin{center}
\includegraphics[width=0.4\textwidth]{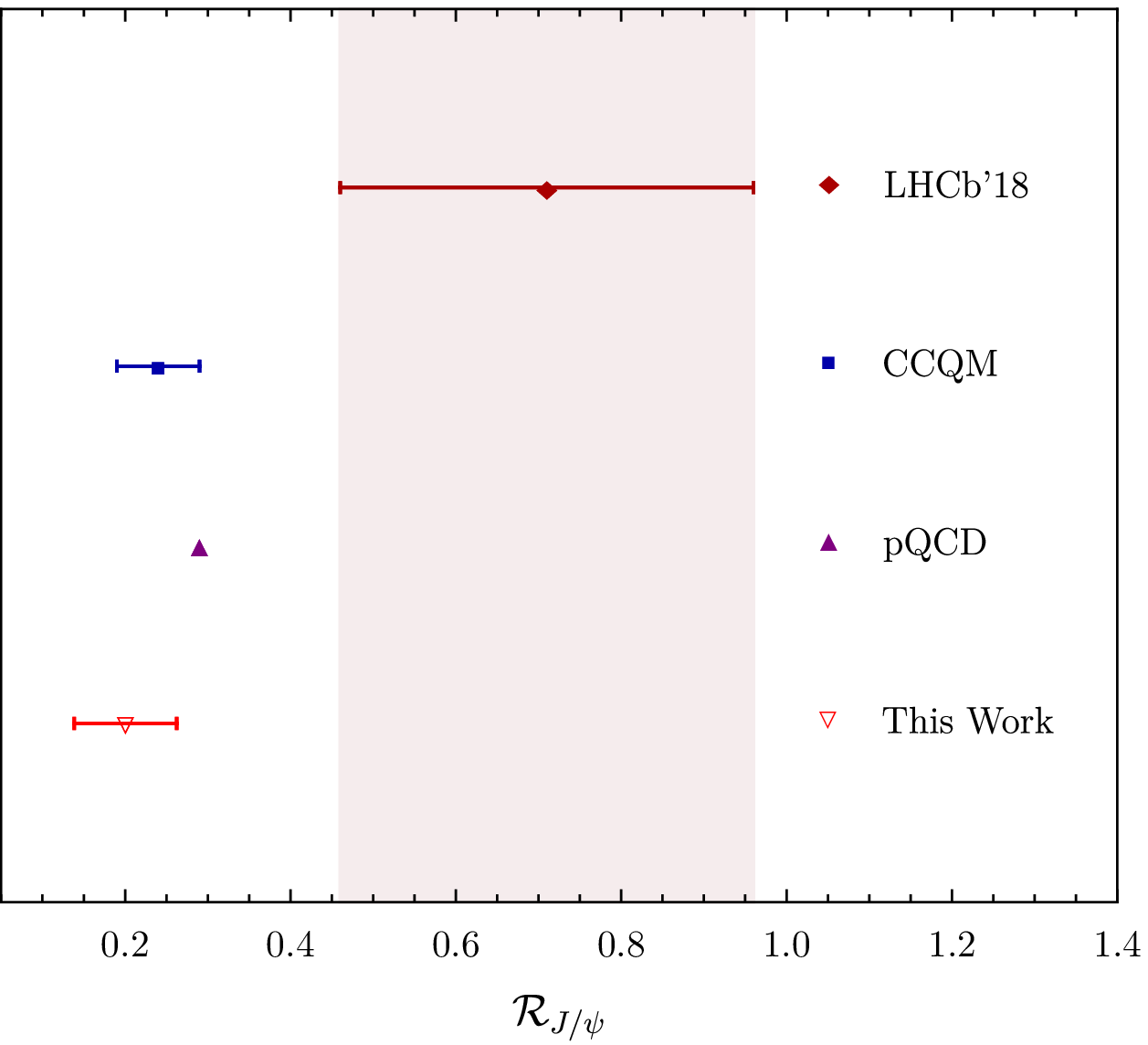}
\includegraphics[width=0.41\textwidth]{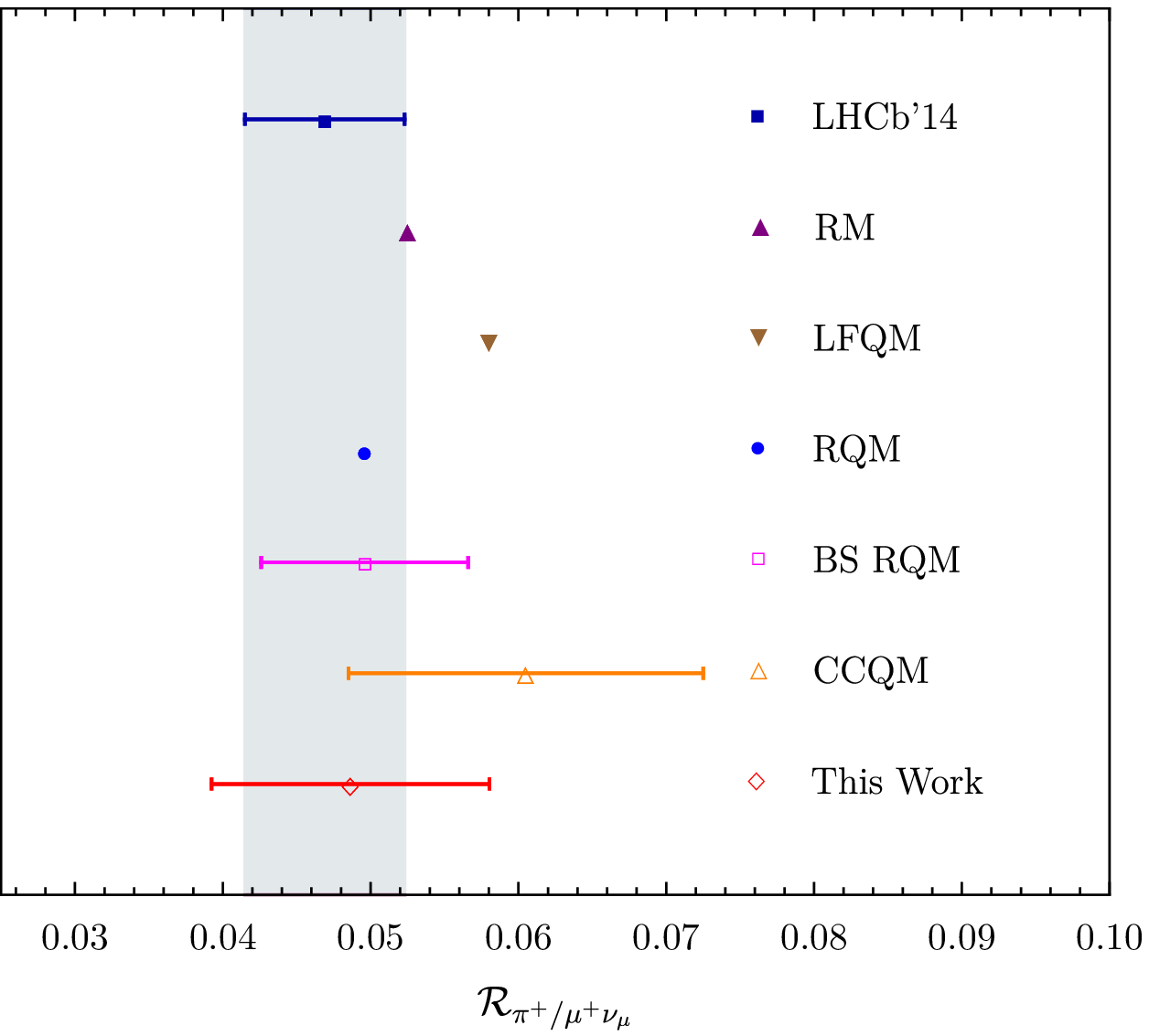}
\end{center}
\caption{The LCSR predictions on branching ratios ${\cal R}_{J/\psi}$ and ${\cal R}_{\pi / \mu^+ \nu_\mu}$ that can be calculated by Eq.~\eqref{eq:Rpimu} and Eq.~\eqref{eq:RJpsi} respectively.}
\label{Fig:R_Jpsi_pimunu}
\end{figure*}

With the $B_c^+ \to J/\psi \ell^+ \nu_\ell$ semi-leptonic decays Eq.~\eqref{Eq:Semi} and $B_c$ lifetime, we can calculate the branching fractions of the semi-leptonic decays $B_c^+ \to J/\psi \ell^+ \nu_\ell$ that are listed in Table.~\ref{tab:BFsemi}.
For comparison, we also present other theoretical predictions, i.e. RM~\cite{AbdElHady:1999xh}, CCQM~\cite{Issadykov:2018myx}, BS RQM~\cite{Chang:2014jca}, RQM~\cite{Ebert:2003cn}, pQCD~\cite{Wen-Fei:2013uea}, LCSR~\cite{Leljak:2019eyw} and QCD-SR~\cite{Azizi:2019aaf}. We find that all the theoretical branching fraction predictions of the $\mu$-lepton decay channel are greater than that of the $\tau$-lepton decay channel. That may be caused by the small mass of $\mu$-lepton. To illustrate this effect more clearly, we have shown the differential decay width in Fig.~\ref{DW:mutaup}, where the solid and shade bands correspond to their central values and the uncertainties respectively. In addition, we can see that even though they were drawn by applying the same formula, however, for $q^2\sim q^2_{\rm min}$, $\mu$- and $\tau$-lepton decay channel are significantly different, and there is an obvious sharp increase for $\mu$-lepton decay channel, which results in the branching fraction of $\mu$ decay channel being significantly larger than that of the $\tau$ decay channel.

Considering that $R_{J/\psi}$ and $R_{\pi/ \mu^+ \nu_\mu}$ were measured by LHCb experiment, their definitions are as follows:
\begin{align}
&{\cal R}_{J/\psi}= \frac { {\cal B}(B_c^+ \to J/\psi \tau^+ \nu_\tau)} {{\cal B}(B_c^+ \to J/\psi \mu^+ \nu_\mu)},
\label{eq:RJpsi}\\
&{\cal R}_{\pi^+/ \mu^+ \nu_\mu}=\frac {{\cal B}(B_c^+ \to J/\psi \pi^+)}{{\cal B}(B_c^+ \to J/\psi \mu^+ \nu_\mu)}. \label{eq:Rpimu}
\end{align}

We collect our results in Table.~\ref{tab:Rall} by using those formula. For comparison, the predictions of LHCb~\cite{Aaij:2014jxa}, LHCb~\cite{Aaij:2016tcz}, LHCb~\cite{Aaij:2017tyk}, RM~\cite{AbdElHady:1999xh}, LFQM~\cite{Ke:2013yka}, CCQM~\cite{Issadykov:2018myx}, BSM~\cite{Chang:1992pt}, BS RQM~\cite{Chang:2014jca}, RQM~\cite{Ebert:2003cn}, pQCD~\cite{Wen-Fei:2013uea}, LCSR~\cite{Leljak:2019eyw}, QCD-SR~\cite{Azizi:2019aaf}, CLFQM~\cite{Huang:2018nnq} are also shown. For convenience, we have shown it in Fig.~\ref{Fig:R_Jpsi_pimunu}. For ${\cal R}_{\pi^+/ \mu^+ \nu_\mu}$, all theoretical predictions are consistent with the LHCb experimental value~\cite{Aaij:2014jxa} within the $2\sigma$-errors. For $R_{J/\psi}$, all theoretical predictions are less than the measured value of the LHCb experiment~\cite{Aaij:2017tyk}. In addition, our predictions of ${\cal R}_{\pi^+/ \mu^+\nu_\mu}$ and ${\cal R}_{K^+/\pi^+}$ are also consistent with other theories, which also shows that the HFFs method is an effective method from the predictions. Therefore, this is a new, complete and effective method for calculating the decay of Bc mesons. Using this method to calculate the ${\cal R}_{J/\psi}$ ratio in SM is also useful.

\section{summary}\label{Section:IV}
In this study, we have investigated the $B_c \to J/\psi$ HFFs and keep up to twist-4 accuracy within the LCSR approach. As shown in Fig.~\ref{Fig:HFF012t}, similar to other pseudoscalar meson decays to vector meson, the $J/\psi$-meson transverse component will not have any contribution to the HFFs of the $B_c \to J/\psi$ transition at the point $q^2=0$. With the extrapolated $B_c \to J/\psi$ HFFs, we study the semi-leptonic decays $B_c \to J/\psi \ell^+ \nu_\ell$ with $\ell=(\mu, \tau)$ and the two-body decays $B_c \to J/\psi + (P, V)$ with $P=(\pi^+, K^+)$ and $V=(\rho^+, K^{*+})$. The corresponding decay widths and branch fractions predictions are listed in Table.~\ref{tab:Rall}. We observe that our results are larger than that of the other theories.

To compare with the experiments, we use these predictions to further study the three kinds of branching ratios, i.e., ${\cal R}_{\pi^+/ \mu^+ \nu_\mu}$ , ${\cal R}_{K^+ /\pi^+}$, and ${\cal R}_{J/\psi}$. The results are listed in Tables~\ref{tab:DwKpi}, \ref{tab:BrBctoKpi}, \ref{tab:BFsemi}. Meanwhile, we also provided two images for comparsion as shown in Figs.~\ref{R:Kpi} and \ref{Fig:R_Jpsi_pimunu} to give our results more clarity. For $\mathcal{R_{\pi^+/\mu^+\nu_\mu}}$ and ${\cal R}_{K^+ /\pi^+}$, our results are consistent with other theorical predictions and the LHCb experimental results within the $1\sigma$-errors. For ${\cal R}_{J/\psi}$, our predictions are close to the those obtained using other theories, but all of the theoretical predictions were smaller than that of the LHCb experimental predictions. Therefore, we believe that the HFFs obtained by the LCSR approach are also applicable to the $B^+_c$ meson two-body decays and semi-leptonic decays $B_c^+ \to J/\psi+(P, V, \ell^+ \nu_\ell)$. According to the ${\cal R}_{J/\psi}$, calculating the HFFs in LCSR in a new way shows that there may be new physics in the $B_c\to J/\psi \ell^+ \nu_\ell$ semi-leptonic decays.

\section*{Acknowledgment}
W. Cheng and H. B. Fu would like to thank the Institute of Theoretical Physics in Chongqing University (CQUITP) for kind hospitality. This work was supported in part by the National Natural Science Foundation of China under Grant No.11765007, No.11625520, No.11947406 and No.12047564, the Project of Guizhou Provincial Department of Science and Technology under Grant No.KY[2019]1171, the Project of Guizhou Provincial Department of Education under Grant No.KY[2021]030, the China Postdoctoral Science Foundation under Grant Nos.2019TQ0329, 2020M670476, the Chongqing Graduate Research and Innovation Foundation under Grant No.ydstd1912, the Fundamental Research Funds for the Central Universities under Grant No.2020CQJQY-Z003, and the Project of Guizhou Minzu University under Grant No. GZMU[2019]YB19.

\end{document}